\documentclass[showpacs,
superscriptaddress,amsmath]{revtex4}

\usepackage{graphicx}
\usepackage{longtable}

\newcommand\ra{\rightarrow}

\newcommand{\lan}{\langle}
\newcommand{\ran}{\rangle}
\newcommand{\ovl}[1]{\overline{#1}}

\newcommand{\non}{\nonumber\\}

\newcommand{\be}{\begin{equation}}
\newcommand{\ee}{\end{equation}}
\newcommand{\bea}{\begin{eqnarray}}
\newcommand{\eea}{\end{eqnarray}}
\newcommand{\ba}[1]{\begin{array}{#1}}
\newcommand{\ea}{\end{array}}

\begin{document}

\title{Productions of hadrons, pentaquarks
$\Theta ^+$ and $\Theta ^{*++}$, and di-baryon
$(\Omega\Omega)_{0^{+}}$ in relativistic heavy ion collisions
by a quark combination model}

 \author{Feng-lan Shao}
 \affiliation{Department of Physics, Shandong University, Jinan,
Shandong 250100, People's Republic of China}
\affiliation{Department of Physics, Qufu Normal University, Qufu,
Shandong 273165, People's Republic of China}
\author{Qu-bing Xie}
\affiliation{China Center of Advanced 
Science and Technology (World Lab),
P.O.Box 8730, Beijing 100080, People's Republic of China}
\affiliation{Department of Physics, Shandong University, Jinan,
Shandong 250100, People's Republic of China}
\author{Qun Wang}
\affiliation{Institut f\"ur Theoretische Physik,
J.W. Goethe-Universit\"at, D-60054 Frankfurt/Main, Germany}
\affiliation{Department of Modern Physics,
University of Science and Technology of China,
Hefei, Anhui 230026, People's Republic of China}

\begin{abstract}

The hadron production in relativistic heavy ion collisions is
well described by the quark combination model. The mixed ratios 
for various hadrons and the transverse momentum spectra for long-life 
hadrons are predicted and agree with recent RHIC data.
The production rates for the pentaquarks 
$\Theta ^+$, $\Theta ^{*++}$ and 
the di-baryon $(\Omega\Omega)_{0^{+}}$ are estimated,   
neglecting the effect from the transition amplitude for 
constituent quarks to form an exotic state. 

\end{abstract}


\maketitle

\section{Introduction}

Recently a lot of data for hadron multiplicities have been
published from the relativistic heavy ion collision (RHIC) at
Brookhaven National Laboratory
\cite{Back:2002wb,Adler:2002sw,Adams:2002pf,Hippolyte:2003yf,
Adler:2002wn,Adcox:2003nr,Back:2002ks,Filimonov:2003xu}.
One of the purpose of this experiment
is to produce a deconfined phase of quarks and gluons (QGP) under
extreme conditions at high temperature and density by smashing two
gold nuclei. Thus it is important to find reliable probes to judge
if the QGP is really formed in the experiment, which is both
theoretical and experimental challenges because one cannot
directly detect the free quarks and gluons but their decay
products. Hadron multiplicities and their correlations are
observables encoding information on chemical properties of the
medium generated from the heavy ion collisions. It is impressive
that the statistical thermal models provide a good description for
the available multiplicity data at almost all energies in heavy
ion collisions by only a few parameters
\cite{Braun-Munzinger:2003zd,Braun-Munzinger:1999qy,Braun-Munzinger:2001ip,
Braun-Munzinger:2003zz,Cleymans:1999st,Koch:1986ud,Rafelski:pu,
Muller:1980kf,Torrieri:2004zz,Torrieri:2003nh,Bratkovskaya:2000qy,
Schenke:2003mj,Rischke:2001bt}.
While the ability to reproduce the mixed particle ratio from
statistical models is not a proof that the emitting source is in
thermal equilibrium \cite{Rischke:2001bt}, 
it is further evidence that such a state,
which is one of the necessary conditions for creation of a QGP,
has been created. The recombination or coalescence models
\cite{Hwa:2002tu, Hwa:2004ng,Greco:2003xt,Greco:2004rm,Greco:2003mm,
Fries:2003kq,Fries:2003vb,Nonaka:2003hx} are very
successful in explaining the RHIC puzzles \cite{Adcox:2001jp,
Adler:2002tq,Adcox:2001mf, Adler:2001bp,Adler:2002pba,Adcox:2002au}.  
e.g. an anomaly in the ratio of proton to pion which is unexpectedly 
high (reaching about 1) at middle transverse momentum region.
This implies the hadronization by 
quark recombination plays an important role even in the hard regime
characterized by high transverse momenta and provides a piece 
of evidence for decomfinement (see, e.g. \cite{Heinz:2004pj}).

The earliest quark combination model (QCM) can be dated back
to 1970s \cite{Anisovich:1972pq,Bjorken:1973mh}, which
was proposed to describe the multiparticle
production or the hadronization in various reactions.
Certainly the most popular hadronization models nowadays are the string
model and the cluster model \cite{Andersson:ia,Webber:1983if,Winter:2003tt},
but the great advantage of quark combination picture in describing
the inclusive hadron production is its simplicity.
The success of the quark combination model in
almost all kinds of high energy collisions,
e.g. electron-positron, hadron-hadron and nucleus-nucleus collisions,
is partly due to the universal stochastic nature of
fragmentation or hadronization. In this sense, the QCM resembles the thermal
or the statistical model, but it encodes more microscopic information.

We have developed a variant of the QCM based on a simple quark
combination rule \cite{Xie:wi,Xie:ap}. Using our QCM, we have
described most of multiplicity data for hadrons in electron-positron and
proton-proton/anti-proton collisions
\cite{Liang:ya,Wang:gx,Wang:ch,Zhao:hq,Wang:dg,Wang:jy,Wang:pg}.
Also we solved a difficulty facing other QCMs in describing the
TASSO data for baryon-antibaryon correlation: they can be
successfully explained by our QCM \cite{Si:ux,Xie:ap}. Embedded in
the event generator, our QCM can also reproduce most of the global
properties of hadronic events such as momentum spectra in
electron-positron collisions
\cite{Si:1999,Wang:1999xz,Wang:2000bv,Li:2002eq,Shao:2003ir}.
Encouraged by the success of the statistical and recombination
models in heavy ion collisions, in this paper we try to extend our
QCM to reproduce the recent RHIC data for hadron multiplicities.
Especially we will predict the production rates of three exotic
baryons: the pentaquarks $\Theta ^+$ and $\Theta ^{*++}$
and di-baryon $(\Omega\Omega)_{0^{+}}$.

The pentaquark $\Theta ^+$ is an exotic baryon made of
five quarks $uudd\overline{s}$, which have been discussed
in the context of quark models in the early days
of QCD \cite{Jaffe:1976ii,Strottman:qu}.
In 1997, Diakonov et. al. \cite{Diakonov:1997mm}
predicted the mass and width of $\Theta ^+$
at about 1540 MeV and 15 MeV respectively,
using the chiral soliton model. Recent works on
the property of the pentaquark in the chiral
field model can be found, e.g. in
Ref.\ \cite{Wu:2003xc,Wu:2003mc,Liu:2004qx}.
Several experimental groups
have reported the discovery
of the pentaquark state $\Theta ^+$
\cite{Nakano:2003qx,Barmin:2003vv,Stepanyan:2003qr,
Airapetian:2003ri,Kubarovsky:2003fi}.
Another pentaquark $\Theta ^{*++}$ belongs to the
27-plet baryon with flavor content $uuud\overline{s}$
and possible spin-parity $J^P=(3/2)^+$ \cite{Wu:2003mc}.
A search for the $\Theta ^{*++}$ pentaquark has been
carried out by BaBar collaboration \cite{Aubert:2004ps} in
the decay channel $B^\pm\ra p\ovl{p}K^\pm$.
But the results in this experiment are preliminary
yet to be confirmed.
The last exotic state we are going to look at is the
di-baryon $(\Omega\Omega)_{0^{+}}$.
For more than 20 years the search for di-baryon
has been another important attempt in hadronic physics
\cite{Jaffe:1977cv,Jaffe:1976yi,Buchmann:wy,
Buchmann:1998mi,Wang:2004nv,Pang:2004mm}.
One believes that if di-baryons do exist
those with multi-strangeness must be
ideal candidates to be observed
because of their relatively long lifetime
due to the stability with respect to strong decay.
Recently the structure and properties of
di-baryons with large strangeness
are investigated in the chiral $SU(3)$ model
\cite{Zhang:ju,Li:2000cb,Yu:2002jm,Zhang:ny}
which quite successfully reproduces several nuclear properties.
They found that some six-quark states
with high strangeness have considerable binding energy
provided by the chiral quark coupling. Of particular interest is
$(\Omega\Omega)_{0^{+}}$: it is a deeply bound state
(binding energy is around 100 MeV; the mean-square root of the distance
between two $\Omega$'s is $0.84$ fm).
The mean lifetime of this di-baryon is
as long as about twice that of $\Omega$, because it only decays weakly.
All these interesting properties together with the electric charge $Q=-2$
would make it easily identified in the experiments.
Because of its large strangeness, $(\Omega\Omega)_{0^{+}}$ is not likely
to be produced in proton-proton collisions.
But one expects the enhanced strangeness production
\cite{Muller:1980kf,Rafelski:pu,Koch:1986ud}
in heavy ion collisions at RHIC energies,
which can be a best place to study the production of
the di-baryon $(\Omega\Omega)_{0^{+}}$ \cite{Pal:2001rp,Zhang:sv}.

The outline of this paper is as follows.
In section \ref{qcr} and \ref{su3} we give a brief
description of the quark combination model and basic relations
among production weights of $SU_f(3)$ multiplets.
In section \ref{hadron} we present
particle ratios and the transverse momentum spectra of the
pion, proton and kaon predicted from the quark combination model.
The upper limits for production rates of $\Theta ^+$ and
$(\Omega\Omega)_{0^{+}}$ are estimated in section \ref{fivequark}
and \ref{diomega}. We give a summary of results and 
conclude in section \ref{summ}.

\section{The quark combination model}
\label{qcr}


All kinds of hadronization models demand themselves, consciously or not,
satisfy rapidity or momentum correlation for quarks in
the neighborhood of phase space. The essence of this correlation is its
agreement with the fundamental requirement of QCD which
uniquely determines the quark combination rule (QCR) \cite{Xie:ap}.
According to QCD, a $q\overline{q}$ may be in a color octet or a singlet
which means a repulsive or an attractive interaction between them.
The smaller the difference in rapidity for two quarks is,
the longer the interaction time is.
So there is enough time for a $q\overline{q}$ to be in a color singlet
and form a meson. Similarly, a $qq$ can be in a sextet or an anti-triplet.
If its nearest neighbor is a $q$ in rapidity, they form a baryon.
If the neighbor is a $\overline{q}$, because the attraction strength
of the singlet is two times that of the anti-triplet,
$q\overline{q}$ will win the competition to
form a meson and leave a $q$ alone. Our quark combination model
is based on the above QCD requirements.
When the transverse momentum of quarks are negligible,
all $q$ and $\overline{q}$ can always line up
stochastically in rapidity. The QCR reads:
\begin{enumerate}
\item Starting from the first parton ($q$ or $\overline{q}$) in the line;

\item If the baryon number of the second in the line is of the different type
from the first, i.e. the first two partons are
either $q\overline{q}$ or $\overline{q}q$,
they combine into a meson and are removed from the line,
go to point 1; Otherwise they are either $qq$ or
$\overline{q}\overline{q}$, go to the next point;

\item Look at the third, if it is of the different type from the first,
the first and third partons form a meson and are removed from the line,
go to point 1; Otherwise the first three partons combine
into a baryon or an anti-baryon and are removed from the line,
go to point 1.
\end{enumerate}

Here is an example to show how the above QCR works:
all partons line up in rapidity and combine to hadrons as follows
\bea
\label{eq1}
&&q_1\ovl{q}_2\ovl{q}_3\ovl{q}_4\ovl{q}_5q_6
\ovl{q}_7q_8q_9q_{10}\ovl{q}_{11}q_{12}q_{13}q_{14}\ovl{q}_{15}
q_{16}q_{17}\ovl{q}_{18}\ovl{q}_{19}\ovl{q}_{20}\non
&&\ra M(q_1\ovl{q}_2)\;\ovl{B}(\ovl{q}_3\ovl{q}_4\ovl{q}_5)\;
M(q_6\ovl{q}_7)\;B(q_8q_9q_{10})\;M(\ovl{q}_{11}q_{12})
\;M(q_{13}\ovl{q}_{15})\;B(q_{14}q_{16}q_{17})\;
\ovl{B}(\ovl{q}_{18}\ovl{q}_{19}\ovl{q}_{20})
\eea
We note that it is straightforward to define
the QCR in 1-dimensional phase space, but it is more complicated to have it
in two or three dimensional phase space, where one does not
have an order or one has to define an order in a sophisticated way so
that all quarks can combine to hadron in a
particular sequence, see for example Ref. \cite{Hofmann:1999jx}.

If the quarks and anti-quarks are stochastically
arranged in rapidity, the probability distribution for
$N$ pairs of quarks and anti-quarks
to combine into $M$ mesons, $B$ baryons and $B$
anti-baryons is
\begin{equation}
\label{eq7}
X_{MB}(N)=\frac{2N(N!)^2(M+2B-1)!}{(2N)!M!(B!)^2}3^{M-1}\delta_{N,M+3B}.
\end{equation}
The average number of primarily produced mesons $M(N)$
and baryons $B(N)$ are given by
\bea
\label{eq8}
\lan M(N) \ran &=&\sum_M\sum_B M X_{MB}(N)\;,\\
\label{eq9}
\lan B(N) \ran &=&\sum_M\sum_B B X_{MB}(N)\;.
\eea
Approximately, for $ N\geq 3$, $\lan M(N)\ran$ and $\lan B(N)\ran$
can be well parameterized as linear functions of quark
number $N$: $\lan M(N)\ran =aN+b$ and
$\lan B(N)\ran =(1-a)N/3-b/3$ where $a=0.66$ and $b=0.56$.
But for $N < 3$,
obviously one has $\lan M(N)\ran =N$ and $\lan B(N)\ran =0$.

Having the number of mesons and baryons in an event, we can obtain
the multiplicity of all primary hadrons from their production weights.
The yield of the hadron $h_i$ can be written as
\begin{equation}
\label{eq14}
\lan h_{i}\ran =\sum_j C_{M_{j}} \lan M\ran \mathrm{Br}(M_j\to h_i)+
\sum_j C_{B_{j}} \lan B\ran \mathrm{Br}(B_j\to h_i)+
\sum_j C_{\ovl{B}_{j}} \lan B\ran \mathrm{Br}(\ovl{B}_j\to h_i),
\end{equation}
where $\lan M \ran$ and $\lan B\ran$ are the average
number of mesons and baryons respectively.
$C_{M_{j}}$, $C_{B_{j}}$ and $C_{\ovl{B}_{j}}$
are the normalized weights for the primary meson $M_{j}$,
the primary baryon $B_{j}$, and the primary anti-baryon $\ovl{B}_{j}$,
respectively. Obviously we have the property $C_{B_{j}}=C_{\ovl{B}_j}$.
$\mathrm{Br}(h_j \to h_i)$ is the weighted branching ratio
for $h_j$ to $h_i$.

The production weights $C_{M_{j}}$ and $C_{B_j}$
for primary hadrons satisfy the $SU_{f}(3)$
symmetry with a strangeness suppression factor $\lambda _s$ for
strange hadrons \cite{Hofmann:1988gy,Casher:wy,Casher:gw}. 
To determine $C_{M_{j}}$ and $C_{B_{j}}$, we need
the ratio $V/P$ of the vector ($J^{P}=1^{-}$)
to the pseudoscalar ($J^{P}=0^{-}$) meson, and the ratio $(3/2)^{+}/(1/2)^{+}$
of the decouplet ($J^{P}=(3/2)^{+}$) to the octet ($J^{P}=(1/2)^{+}$)
baryon. Assuming $SU(6)$ symmetry, Anisovich et al. \cite{Anisovich:1972pq}
gave $V/P=3$. In this case the weights for all mesons except
$\eta$ and $\eta'$ can be simply written as
$C_{M_{i}} \propto (2J_{i}+1) \lambda_s^{r_{i}}$
where $J_{i}$ is the spin of $M_{i}$, and $r_{i}$
is the number of strange quarks and/or anti-quarks in the meson.
However there is a spin suppression effect associated with
baryons. In the next section, we derive the
two relations between the production weights for octet, decuplet and
singlet baryons from the properties of hadronization in the quark
combination scheme.

\section{$SU_f(3)$ symmetry and flavor conservation
in the quark combination model}
\label{su3}

Hadronization is the soft process of the strong interaction and is
independent of flavor, so the net flavor number remains constant
during the process, which we call the property of flavor
conservation. In the quark combination scheme, this means that the
number of quarks of a certain flavor prior to hadronization equals
to that of all primarily produced hadrons after it. The
$\lambda_s$-broken $SU_f(3)$ symmetry in hadron production means
baryons or mesons in the same $J^{PC}$ multiplet share an equal
production rate up to a $\lambda_s^r$ factor. This $SU_f(3)$
symmetry has been supported by many experiments, particularly by
the fact that the observed $\lambda_s$ obtained from various
mesons and baryons coincide with each other \cite{Hofmann:1988gy}.
This experimental fact is unexpected in the usual diquark model
for baryon production and turns out to be in favor of the quark
combination scheme \cite{Scheck} where a baryon is formed by the
stochastic combination of three constituent quarks.

\begingroup
\squeezetable
\begin{table}
\label{weight}
\caption{The weights of quark combination and those
in terms of baryon multiplets.}
\tabcolsep0.3in
\arrayrulewidth1pt
\begin{tabular}{|c|c|c|}
\hline
Flavor content & Combination weight & Production weight of baryon multiplets \\
\hline
$uuu$  & 1 & $p_{10}$\\
\hline
$ddd$  & 1 & $p_{10}$\\
\hline
$sss$  & $\lambda^3$  & $p_{10}\lambda^3$\\
\hline
$uud$ & 3 & $p_{10}+P_{8}$\\
\hline
$uus$ & $3\lambda$  & $[p_{10}+p_{8}]\lambda$\\
\hline
$ddu$ & 3 & $ p_{10}+P_{8}$\\
\hline
$dds$ & $3\lambda$  & $[p_{10}+p_{8}]\lambda$\\
\hline
$uss$ & $3\lambda^2$ & $[p_{10}+p_{8}]\lambda^2$\\
\hline
$dss$ & $3\lambda^2$ & $[p_{10}+p_{8}]\lambda^2$\\
\hline
$uds$ & $6\lambda$  & $[p_{10}+2(p_{8}+P_{1'})]\lambda$\\
\hline
\end{tabular}
\end{table}
\endgroup

From the above properties, we can obtain relations among the
production weights for octet, decuplet and singlet baryons in the
quark combination picture\cite{Wang:ch}. As shown in Table I, all
the flavor combinations are listed in the first column, with their
combination weights in the second. In hadronization three quarks
combine into a primary baryon which satisfies the
$\lambda_s$-broken $SU_f(3)$ symmetry. As each flavor combination
in the first column corresponds to a certain baryon belonging to
several $SU_f(3)$ multiplets whose production weight is listed in
the third column. The ground-state decuplet and octet baryons are
denoted by 10 and 8, while the only excited baryon which we
consider, the singlet $\Lambda(1520)$, is denoted by $1'$. Their
corresponding weight are denoted by $P_{10}$, $P_8$, and $P_{1'}$
respectively. Note that the two sets of weights must be associated
with a common factor, we finally derive the following relations
among the production weights for octet, decuplet and singlet
baryons: \bea \label{eq13a}
P_{10}&=&P_{1'}\\
\label{eq13b}
P_8&=&2P_{10}
\eea
They impose a global constraint on the production rates of all ground
state baryons and the excited singlet baryon $\Lambda(1520)$.
We can understand the so-called spin suppression from the basic relations.
For those decuplet and octet baryons which are primarily produced
and have the same strangeness, the ratio of their production rate is
\begin{equation}
\label{eq14a}
R=\frac{(3/2)^{+}}{(1/2)^{+}}=\frac{P_{10}}{P_{8}}=0.5.
\end{equation}
Note that the ratio $R$ measured in experiments
is for the octet baryons which include decay
products from the decuplet ones, we then obtain
the following approximated value for $R_{exp}$
when neglecting the production of excited baryons:
\begin{equation}
\label{eq14b}
R_{exp}=\frac{(3/2)^{+}(\mathrm{ground})}{(1/2)^{+}(\mathrm{ground})}
\sim \frac{P_{10}}{P_{10}+P_{8}}\sim 0.3 .
\end{equation}
Note that $R_{exp}$ is much less than 2 from spin counting, which is
called the spin suppression effect.

We can see that there is an essential difference
between the spin suppression factor $R$ for the baryon and
the multiplicity ratio of vector mesons to pseudo-scalar mesons $V/P$.
The former is for the ratio of
the $(3/2)^+$ decuplet baryon to the $(1/2)^+$ octet one.
They belong to different $SU_f(3)$ multiplets.
The basic relations impose a
constraint upon their production weights. The latter is the ratio of
vector to pseudo-scalar mesons, both of which belong to $SU_f(3)$ nonets.
Flavor conservation and $SU_f(3)$ symmetry hold in each nonet and
hence there is no restriction on their weights.

\section{Hadron multiplicities and momentum spectra}
\label{hadron}

In this section,we use the QCM to 
compute hadron multiplicities and their 
ratios in heavy ion collisions at RHIC
energies. We will also calculate the transverse 
momentum spectra for pions, kaons and protons. 
Before we do that, we have to determine some input
parameters of the QCM. The parameters which control the total
multiplicity are the number of quarks and anti-quarks. In
electron-positron and proton-antiproton collisions, the number of
quarks is equal to that of anti-quarks, which means there are no
excess baryons in contrast to anti-baryons. But for
nucleus-nucleus collisions, we need two parameters, the number of
quarks and that of anti-quarks, to account for the net baryon
number even in central rapidity. These two parameters are
determined by fitting the the total charged multiplicity data,
$\langle N_{ch}\rangle _{data}=4100\pm{210}$, in central Au+Au
collisions at 130 AGeV \cite{Back:2002wb}, which
corresponds to the total number of quarks and anti-quarks
$\langle N_{q}+N_{\ovl{q}}\rangle =7400$ in the QCM.
The quark number $\langle N_q\rangle$ and the anti-quark number $\langle
N_{\ovl{q}}\rangle$ can be further determined by the ratio of
antiproton to proton $\ovl{p}/p=0.7$
\cite{Adams:2002pf,Hippolyte:2003yf}. Then we obtain $\langle
N_q\rangle = 3920$ and $\langle N_{\ovl{q}}\rangle = 3500$, where
we see that the net quark number is about 420. Another parameter
is the strangeness suppression factor $\lambda _s$ which has also
to be input from data. We find $\lambda _s=0.5$ is consistent with data
$\Phi/K^{*0}=0.47$ \cite{Adler:2002sw}.
At 200 AGeV, we determine in the same way the total number of quarks and
anti-quarks $\langle N_{q}+N_{\ovl{q}}\rangle =8900$
with the net quark number 360 and $\lambda _s=0.6$ by fitting the data 
\cite{Nouicer:2002ks}.

\begingroup
\squeezetable
\begin{table}
\label{ap2p}
\caption{Anti-particle to particle ratios: our predictions and STAR data at
130 AGeV \cite{Adams:2002pf} and 200 AGeV \cite{Hippolyte:2003yf}.
The second and fourth columns are STAR data at 130 and 200 AGeV respectively.
The third and fifth columns are our results
at 130 and 200 AGeV respectively.}
\tabcolsep0.4in
\arrayrulewidth1pt
\begin{tabular}{|c|c|c|c|c|}
\hline
 & STAR (130) & QCM & STAR (200) & QCM\\
\hline
$\ovl{p}/p$  & 0.71$\pm$0.01$\pm$0.04 & 0.71
& 0.73 $\pm$ 0.05 & 0.78\\
\hline
$\ovl{\Lambda}/\Lambda$  & 0.71$\pm$0.01$\pm$0.04 & 0.79
& 0.84 $\pm$ 0.05 & 0.84 \\
\hline
$\ovl{\Xi}^{+}/\Xi^{-}$  & 0.83$\pm$0.04$\pm$0.05 &0.88
& 0.94$\pm$0.08 & 0.91 \\
\hline
$\ovl{\Omega}^{+}/\Omega^{-}$ & $0.95\pm{0.15}\pm{0.05}$ & 1.00
& $1.03\pm{0.12}$ & 1.00\\
\hline
\end{tabular}
\end{table}
\endgroup

\begin{table}
\label{kp2kn}
\caption{$K^{+}/K^{-}$ ratio compared with compound ratios of
baryons at 130 AGeV \cite{Adams:2002pf}.}
\tabcolsep0.4in
\arrayrulewidth1pt
\begin{tabular}{|c|c|c|}
\hline
 & STAR & QCM \\
\hline
$K^{+}/K^{-}$  & $1.092\pm{0.023}$ & 1.122\\
\hline
$\frac{\ovl{\Lambda}/\Lambda}{\ovl{p}/p}$ & $0.98\pm{0.09}$ & 1.10\\
\hline
$\frac{\ovl{\Xi}^{+}/\Xi^{-}}{\ovl{\Lambda}/\Lambda}$ & $1.17\pm{0.11}$ &1.11 \\
\hline
$\frac{\ovl{\Omega}^{+}/\Omega^{-}}{\ovl{\Xi}^{+}/\Xi^{-}}$ & $1.14\pm{0.21}$ & 1.15\\
\hline
\end{tabular}
\end{table}

\begingroup
\squeezetable
\begin{table}
\label{sp2p}
\caption{A group of ratios for strange hadrons
compared with STAR data at 130 AGeV
\cite{Adams:2002pf,Hippolyte:2003yf,Adler:2002wn}.}
\tabcolsep0.15in
\arrayrulewidth1pt
\begin{tabular}{|c|c|c|}
\hline
 & DATA & QCM \\
\hline
$\Phi/K^{*0}$ & $0.49\pm{0.05}\pm{0.12}$ & 0.47 \\
\hline
$K^{+}/\pi^{-}$ & $0.161\pm{0.002}\pm{0.024}$ & 0.146\\
\hline
$K^{-}/\pi^{-}$ & $0.146\pm{0.002}\pm{0.022}$ & 0.130\\
\hline
$(\ovl{\Omega}^{+}+\Omega^{-})/h^{-}$
& $(2.24\pm{0.69})\times 10^{-3}$ & 3.21$\times 10^{-3}$\\
\hline
\end{tabular}
\end{table}
\endgroup

Having determined the above parameters, we calculate the ratios of
strange anti-baryons to their baryon counterparts,
$\ovl{\Lambda}/\Lambda$,
$\ovl{\Xi}^{+}/\Xi^{-}$,$\ovl{\Omega}^{+}/\Omega^{-}$, in
mid-rapidity of central Au+Au collisions and compare our results
with STAR data \cite{Adams:2002pf,Hippolyte:2003yf}. The above
antibaryon to baryon ratios are mainly controlled by the net quark
number from colliding nucleons.
The results are listed in Table II.
The data show that the ratios increase with strangeness of the
baryons. The reason for this trend is that the quark pair
production is more important than the baryon transport in mid-rapidity.
We see that there is a good agreement
between our model predictions and the datak, which means the quark
combination mechanism can explain this behavior. In the QCM various
ratios $\ovl{B}/B$ are associated with a common multiplicative
factor $D$ which is given by the ratio $K^{+}/K^{-}$, for example
$\ovl{\Lambda}/\Lambda =D (\ovl{p}/p)$,
$\ovl{\Xi}/\Xi=D(\ovl{\Lambda}/\Lambda)$ and $\ovl{\Omega}/\Omega
= D(\ovl{\Xi}/\Xi)$. The calculated ratio $K^{+}/K^{-}$
and the $D$-factor in three compound ratios as shown above.
The results are listed in Table III.
We also calculate the ratios of singly-strange particles to
non-strange particles and those of multi-strange particles to
singly-strange particles. The results and the experimental data
\cite{Adams:2002pf,Hippolyte:2003yf,Adler:2002wn} are given in
Table IV and V. All above show that the mixed ratios for various hadrons
agree with RHIC data.


\begingroup
\squeezetable
\begin{table}
\label{ratio200GeV}
\caption{A group of ratios for strange hadrons
compared with STAR data at 200 AGeV \cite{Back:2002ks,Filimonov:2003xu}.}
\tabcolsep0.15in
\arrayrulewidth1pt
\begin{tabular}{|c|c|c|}
\hline
 & DATA & QCM \\
\hline
$K^{-}/\pi^{-}$ & $0.146\pm{0.024}$ & 0.132\\
\hline
$\ovl{p}/\pi^{-}$ & $0.09\pm{0.01}$ & 0.089\\
\hline
$\Omega^{-}/h^{-}$ & $(1.20\pm{0.12})
\times 10^{-3}$ & $1.40\times 10^{-3}$\\
\hline
$\pi^{-}/\pi^{+}$ & $1.025\pm{0.006}\pm{0.018}$ & 1.008\\
\hline
$K^{-}/K^{+}$ & $0.95\pm{0.03}\pm{0.03}$ & 0.92\\
\hline
\end{tabular}
\end{table}
\endgroup

\begin{table}
\label{pentaquark}
\caption{Upper limits for multiplicities 
of $\Theta^+$, $\Theta^{*++}$, and
$(\Omega\Omega)_{0^{+}}$ at 130 AGeV and 200 AGeV.
The second and third columns are
predictions at 130 AGeV, while the fourth and fifth columns are
predictions at 200 AGeV. Per event values are 
obtained by taking the full rapidity range as $|y|<3.43$. }
\tabcolsep0.3in
\arrayrulewidth1pt
\begin{tabular}{|c|c|c|c|c|}
\hline
 & per rapidity (130) & per event & per rapidity (200) & per event\\
\hline
$\Theta^+$ & 1.16 &  7.97 & 1.30 & 8.93\\
\hline
$\Theta^{*++}$ & 1.86 & 12.75 & 2.08  & 14.29\\
\hline
$\Omega^-$ & 0.40 & 2.74 & 0.51 & 3.50 \\
\hline
$(\Omega\Omega)_{0^+}$ & 2.70$\times 10^{-5}$ & 1.86$\times
10^{-4}$ & 3.67$\times 10^{-5}$ & 2.52$\times 10^{-4}$ \\
\hline
$(\Omega\Omega)_{0^+}/\Omega^{-}$ &6.79$\times 10^{-5}$
& 6.79$\times 10^{-5}$ & 7.20$\times 10^{-5}$ &
 7.20$\times 10^{-5}$\\
\hline
\end{tabular}
\end{table}

As a test of our QCM, we finally calculate the transverse momentum
spectra for the pion, kaon and proton.  
Two steps are needed to get final hadron spectra. 
One is to obtain the spectra for initially produced hadrons 
which are those before their decay. We can do this through our Quark
Combination Model. The next step is to let these initially produced 
hadron decay to final state hadrons which are observed in experiments. 
In our QCM, the longitudinal momentum is described by the 
constant distribution of rapidity, where we assume that quarks 
and anti-quarks are distributed in rapidity range $y\in [-2.8,2.8]$ 
at 130 GeV and $y\in [-3.0,3.0]$ at 200 GeV with equal probability. 
This corresponds to the rapidity distribution of charged 
multiplicity at RHIC.
Once the transverse momentum and rapidity for each quark are 
determined, its momentum is fixed. We can randomly line up 
all quarks in rapidity and let them combine into hadrons following 
our combination rule in section \ref{qcr}. The momentum of 
a hadron is then the sum of that of its constituent quarks, while 
its energy is given by the mass-shell condition. 
The transverse momentum distribution of quark or anti-quark 
is extracted from the measured neutral pion spectra at 200 AGeV 
\cite{Adler:2003qi}, where we assume that they are 
identical because the observed ratios 
of anti-particles to particles, i.e.
$\ovl{p}/p$, $\ovl{\Lambda}$/$\Lambda$, $\ovl{\Xi}^{+}$/$\Xi^{-}$
and $K^{+}/K^{-}$, are almost constant with $p_T$ \cite{Adams:2002pf}. 
This is what we get for quark or antiquark $p_T$ distribution: 
$f(p_T)=(p_T^{2.3}+p_T^{0.2}+1)^{-3.0}$. 
We assume that the combination occurs
only among those constituent quarks which have the same azimuthal
angle, similar to the strategy in Ref.\ \cite{Hwa:2002tu}. 
Note that all these hadrons after combination are primarily
produced, in order to get the spectra comparable to data 
one has to let them decay in their center-of-mass system. 
The momenta of all decayed hadrons are then boosted back
into laboratory frame and recorded. Here we make use of 
the approriate function of event generator JETSET. 
Finally we obtain the $p_T$ spectra for the pions, kaons 
and protons in final state which includes all 
contributions from decay.

That we distinguish primarily produced hadrons 
from the final state ones marks the
{\it essential difference} between our model 
and other recombination or coelescence models 
\cite{Hwa:2002tu,Hwa:2004ng,Greco:2003xt,Greco:2004rm,Greco:2003mm,
Fries:2003kq,Fries:2003vb,Nonaka:2003hx} 
where one does not. Just to get the feeling of
this substantial difference, one may take pions as an example: 
even in the range $p_T>2GeV/c$, about 70\% of pions come 
from resonance decays. This is the reason why 
our quark $p_T$ spectrum extracted 
from data decreases not exponentially at low transverse momenta but 
in a power law -- {\it it reflects the spectra of 
initially produced hadrons instead of final state ones}. 
Another difference between our model and other combination 
models is that the quark $p_T$ distribution is given 
by one single function without distinguishing 
thermal quarks (low $p_T$) and shower quarks (high $p_T$).  
In any combination models the role of gluons has been 
finally taken by quark-antiquark pairs, for example, 
shower quarks are developed after a QCD cascade. 
In our QCM, the number of constituent quarks already 
effectively includes those quark-antiquark pairs converting from gluons. 
All quarks in the whole $p_T$ range are assumed to 
combine following the same combination rule. 
This assumption is widely used in other 
combination models: combination among shower quarks 
is treated in the same footing as among shower quarks 
and thermal quarks. The difference just lies in the 
sources: shower quarks are products after QCD showering 
while thermal quarks are from thermal distribution. 
Except these differences, our model is 
basically the same as other combination models. 
This can be seen that the rapidity correlation principle 
in our model is also implied in other models: in Hwa-Yang's model, 
they assumed $y_1=y_2=y_h$ and $(y_1+y_2)/2=y_h$ where 
$y_1$ and $y_2$ are quark rapidities, and $y_h$ 
is the hadron one \cite{Hwa:2002tu}; While in GKL model 
\cite{Greco:2003xt}, they used even more simple 
assumption $y_1=y_2=y_h=0$.

\begin{figure}
\includegraphics[scale=0.4]{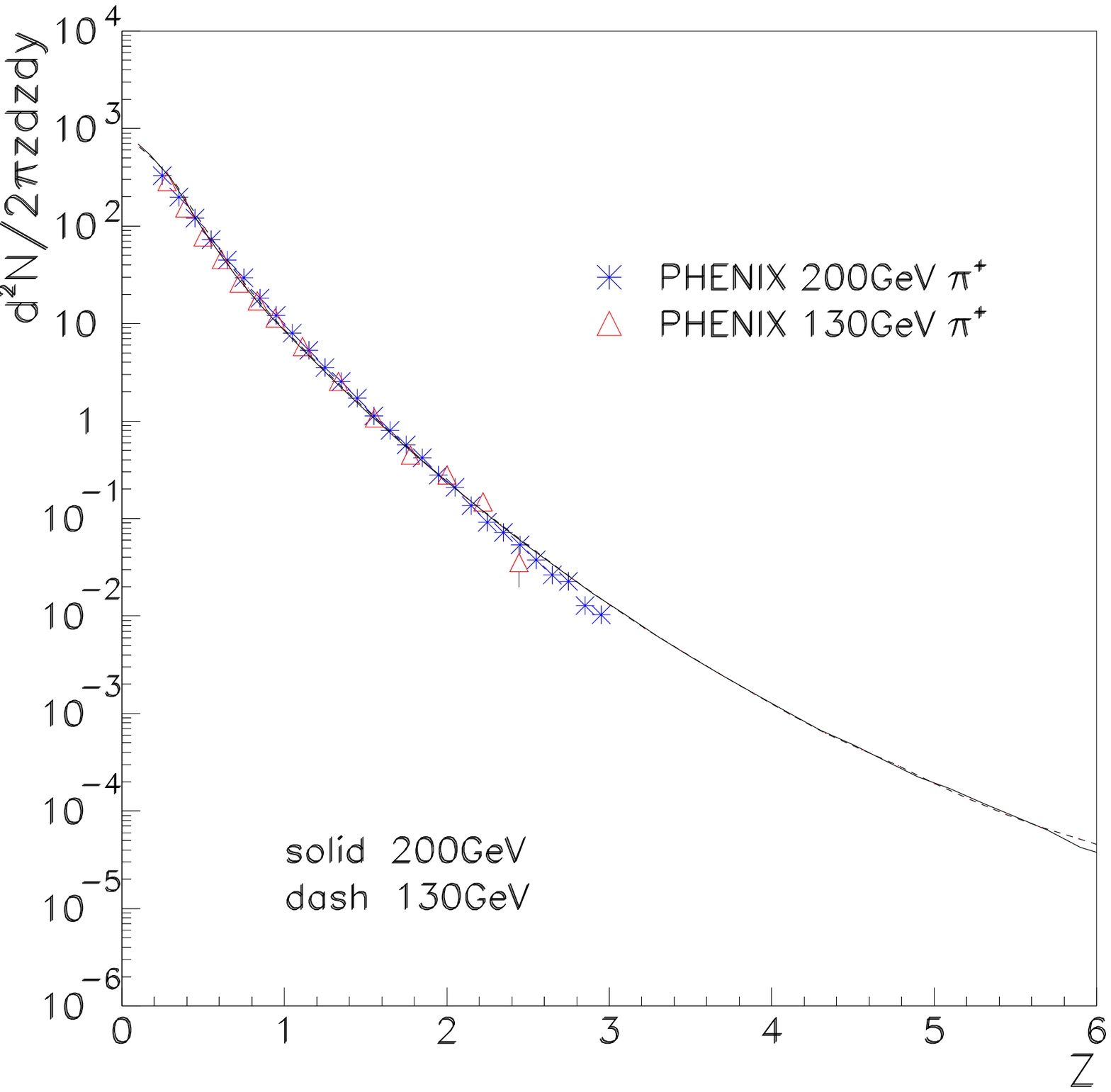}
\includegraphics[scale=0.4]{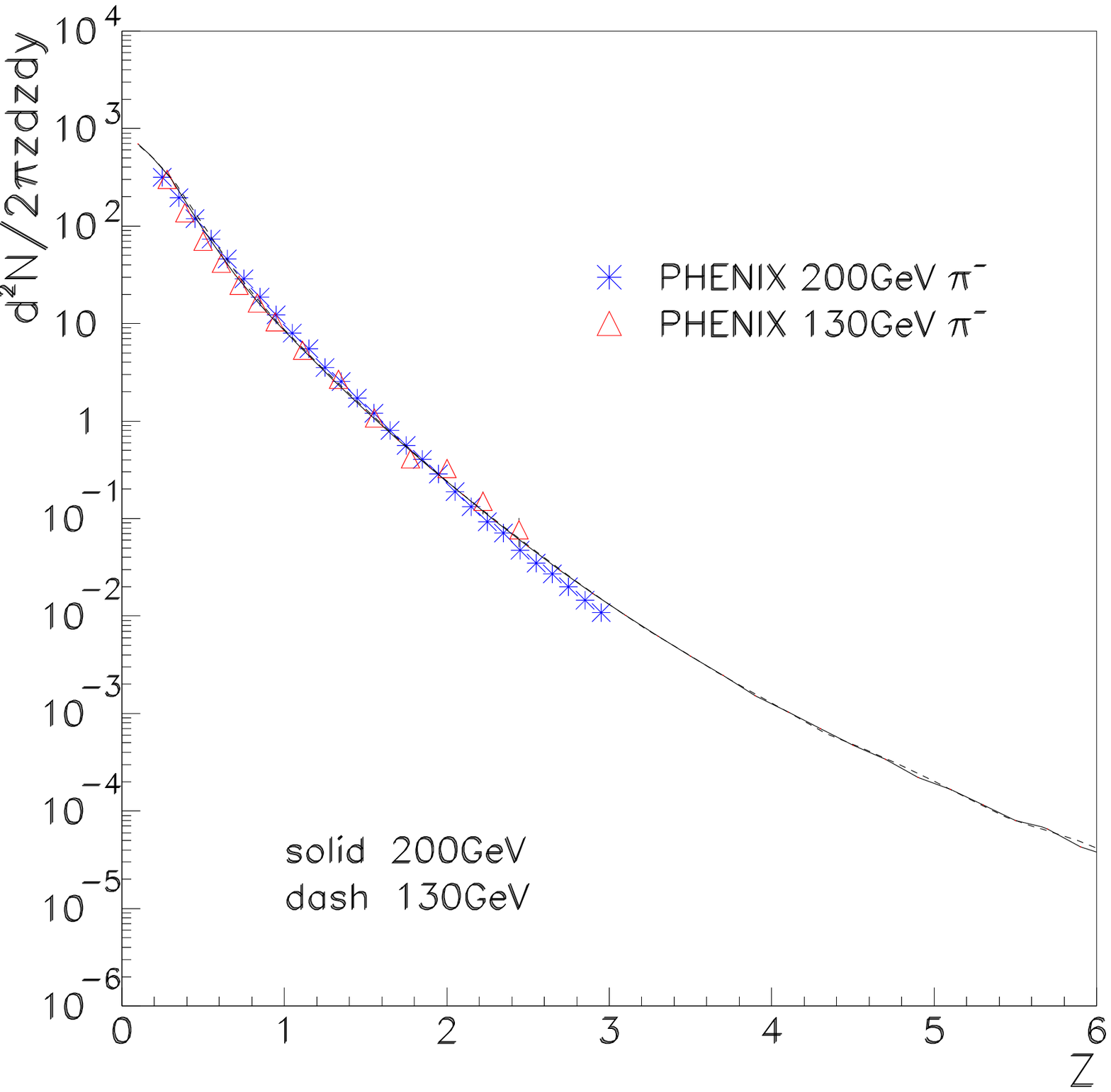}
\includegraphics[scale=0.4]{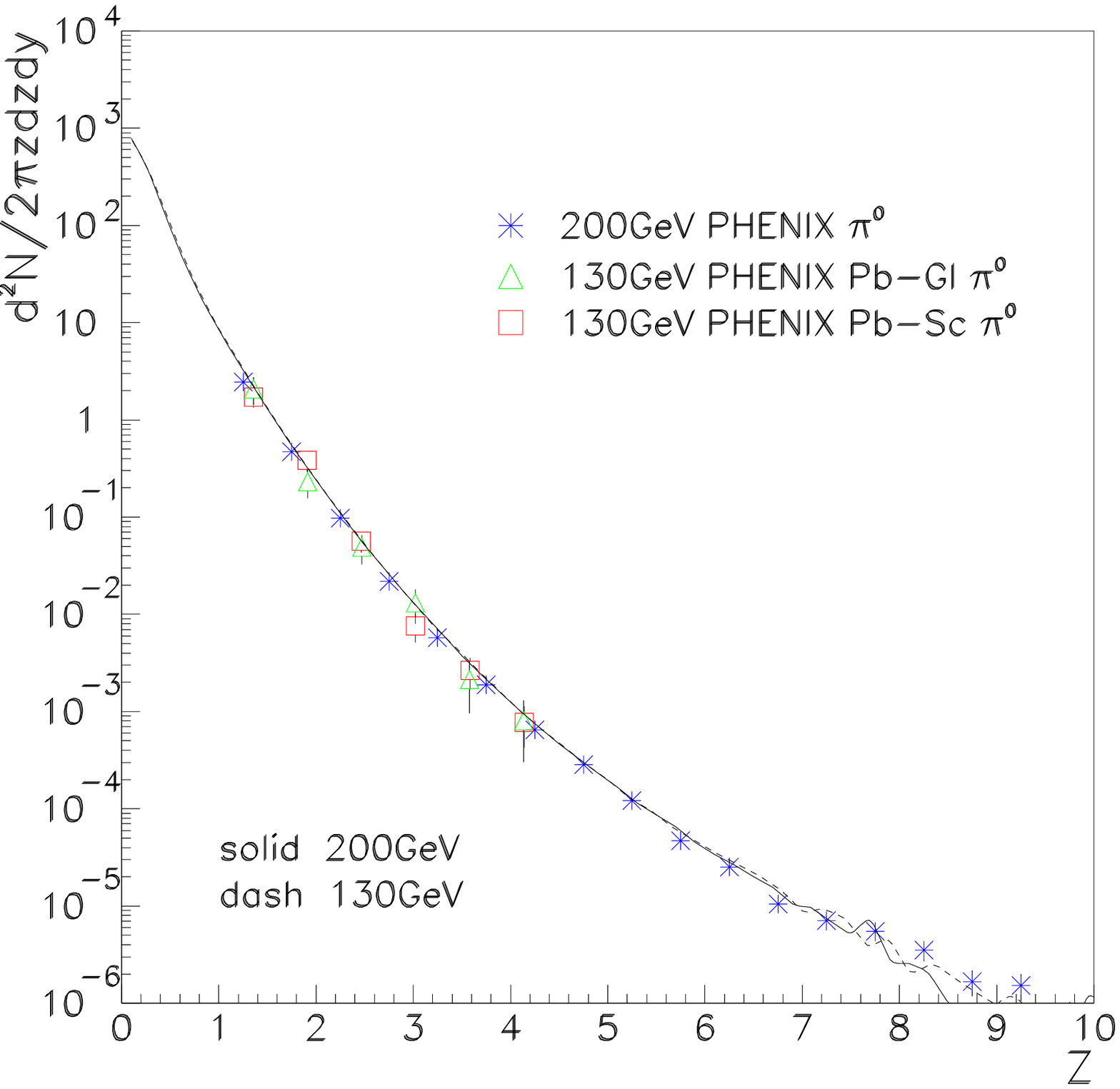}
\caption{
The spectra in $p_T$ scaling variable $z=p_T/K$ ($K$ is the scaling 
factor) for $\pi^\pm$ and $\pi^0$ in the $5\%$ and $10\%$ 
most central collisions at 130 and 200 AGeV, respectively. 
The solid and dashed lines are our results. 
The data are from PHENIX.}
\label{Fig.1}
\end{figure}

\begin{figure}
\includegraphics[scale=0.4]{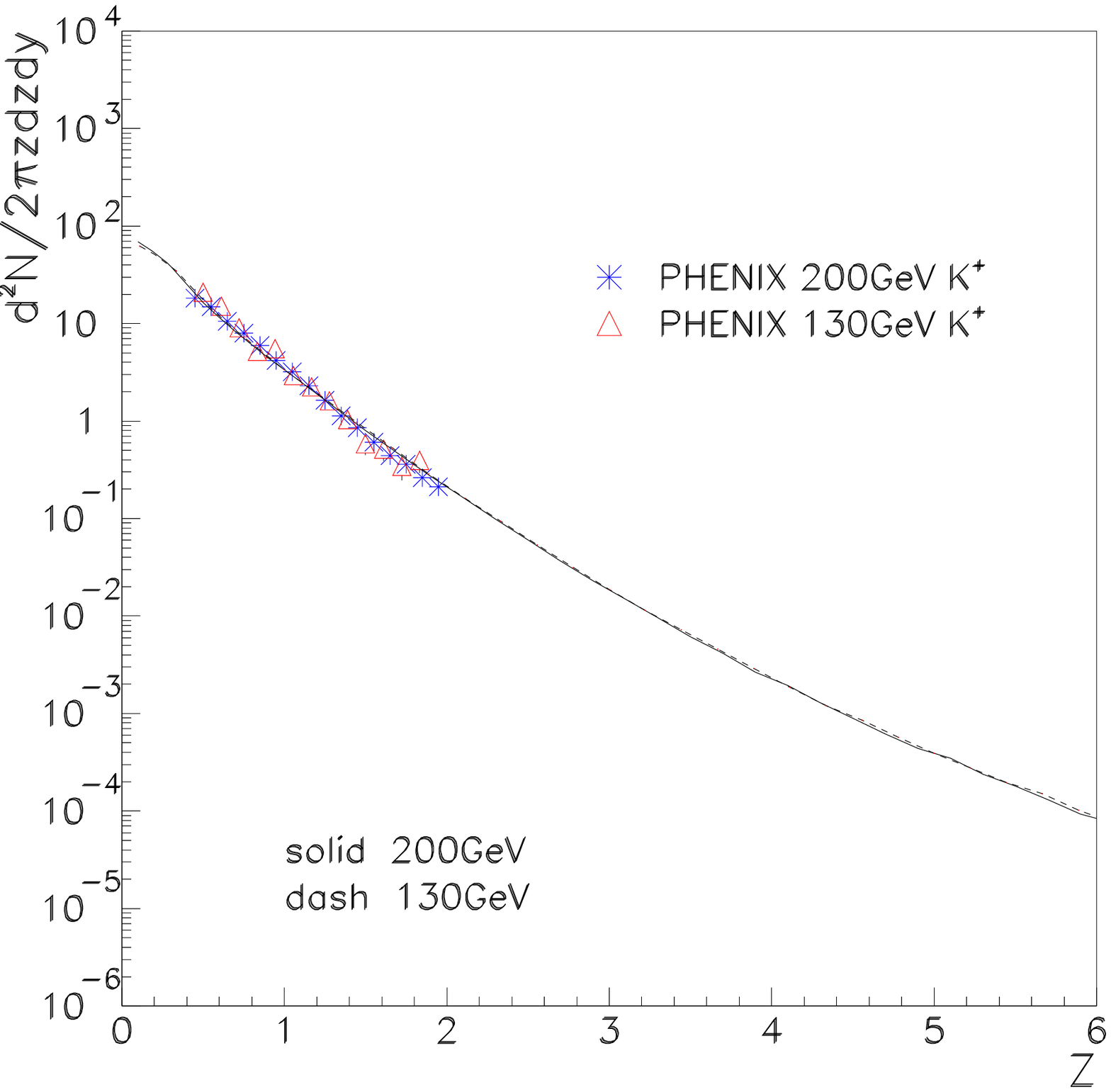}
\includegraphics[scale=0.4]{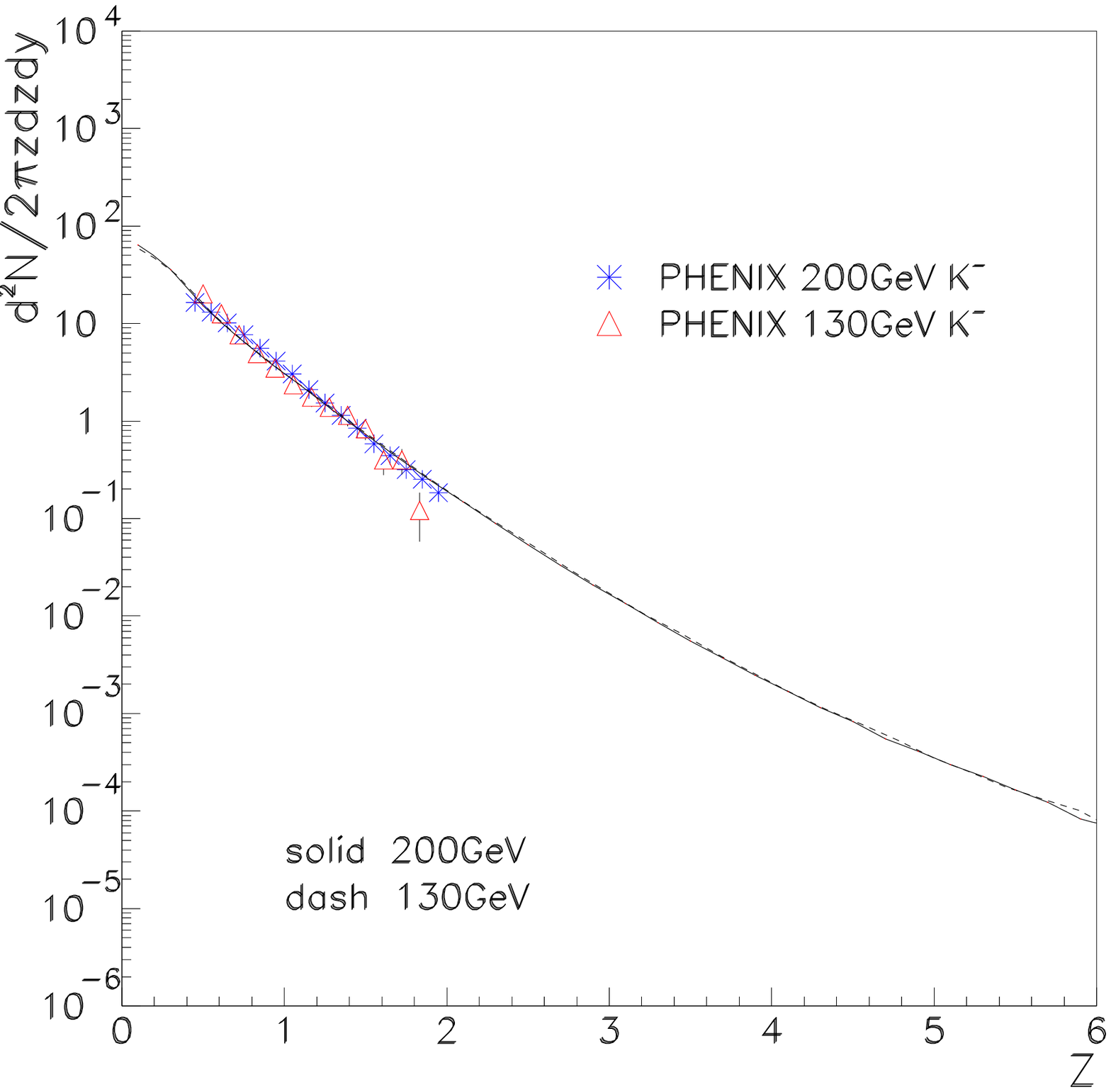}
\caption{
The same spectra for $K^\pm$ in the $5\%$ most 
central collisions at 130 and 200 AGeV.
The solid and dashed lines are our results.
The data are from PHENIX. }
\label{Fig.2}
\end{figure}

\begin{figure}
\includegraphics[scale=0.4]{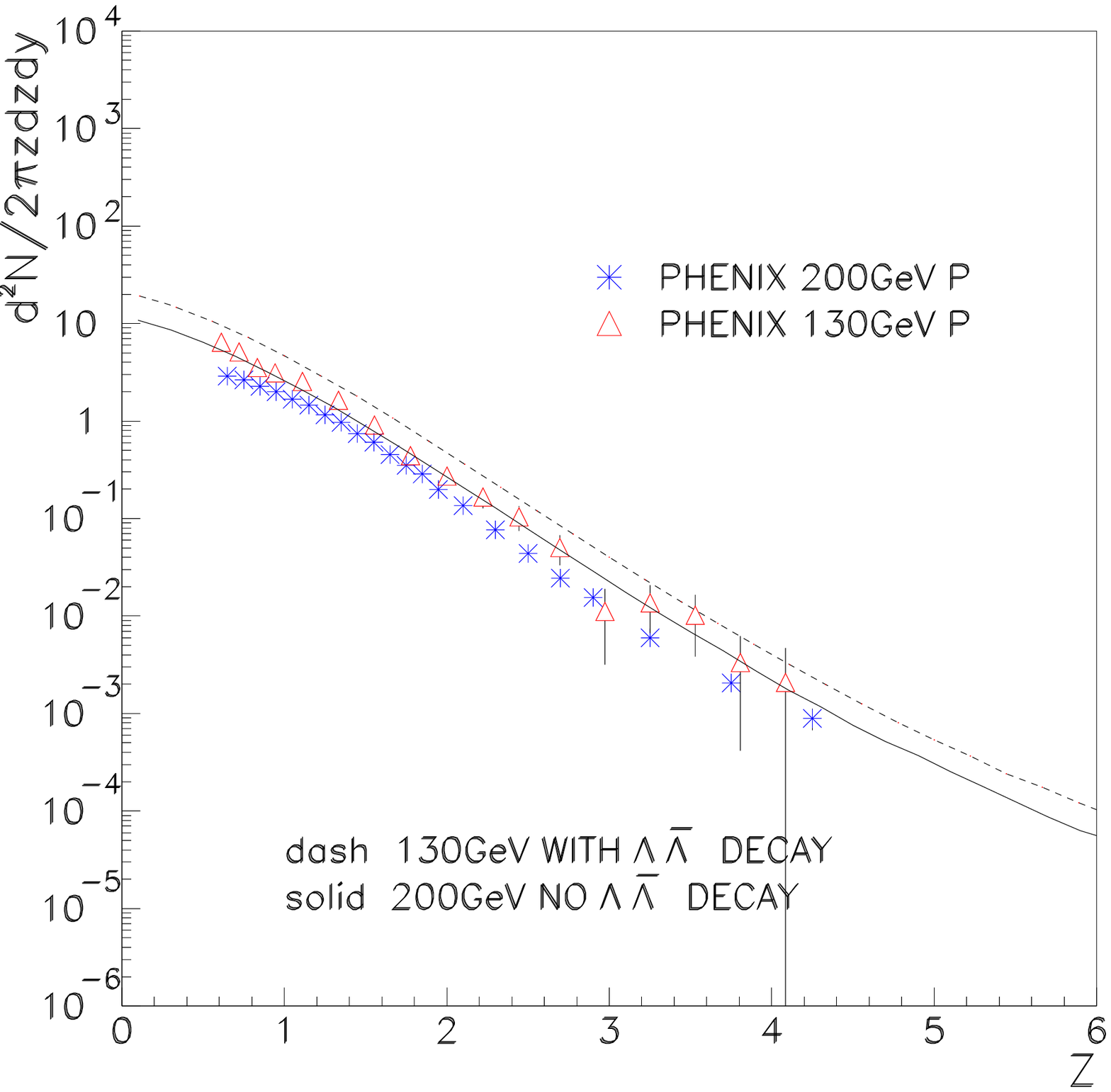}
\includegraphics[scale=0.4]{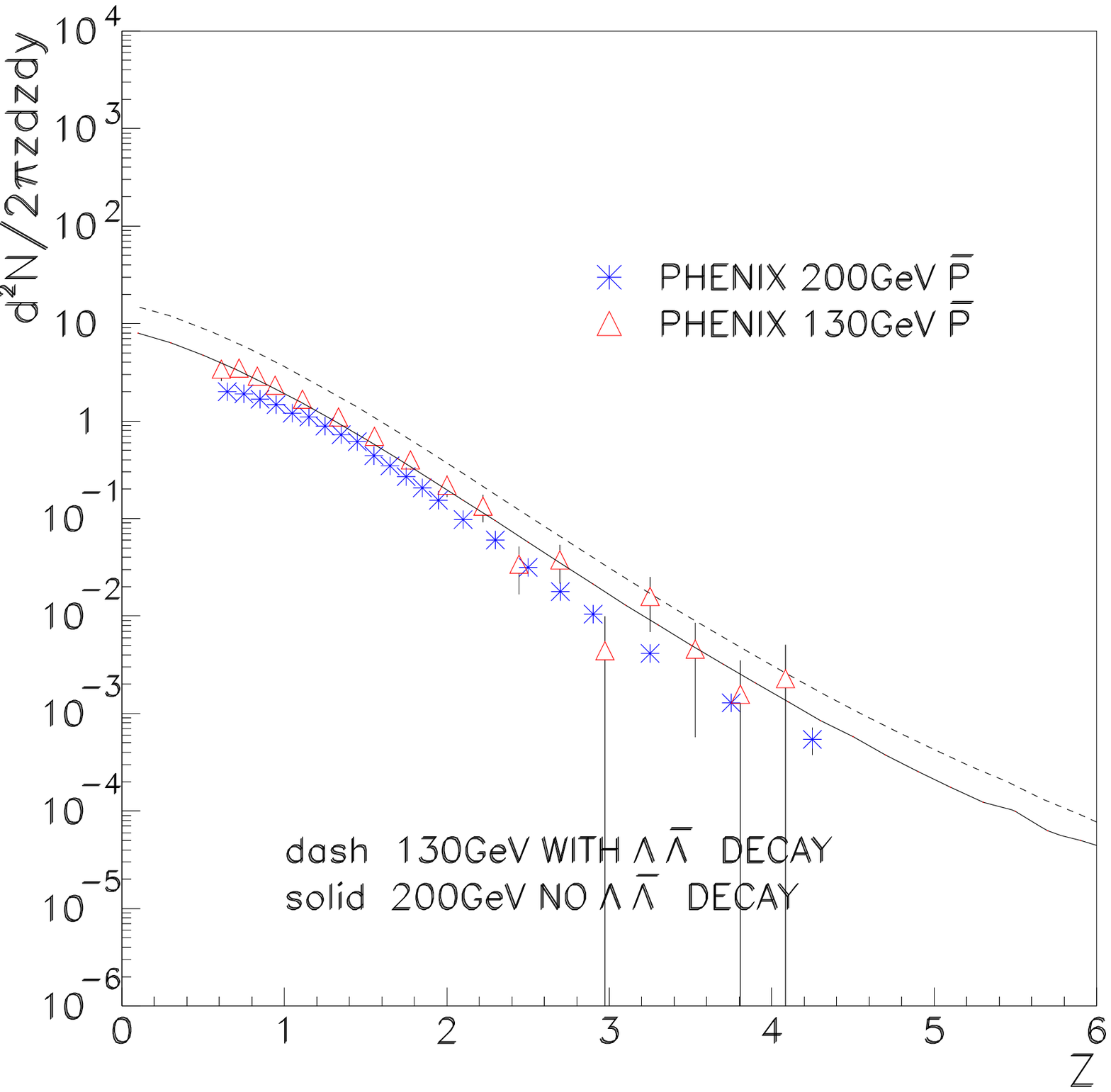}
\caption{
The same spectra for $p,\overline{p}$ in the $5\%$ most
central collisions at 130 and 200 AGeV.
The solid and dashed lines are our results.
The data are from PHENIX. }
\label{Fig.3}
\end{figure}

Instead of giving the prediction for $p_T$ spectra, 
we give the spectra of the scaling variable $z=p_T/K$ where 
the scaling factor $K$ is defined in Ref.\cite{Hwa:2002tu}.   
The result for pions, kaons and protons at 130 and 200 AGeV 
are shown in Fig. (\ref{Fig.1}-\ref{Fig.3}). We see that the agreement 
between our predictions and data is quite satisfactory and 
we have verified the perfect scaling behavior in our model. 
The observed enhancement of protons and antiprotons at
intermediate transverse momenta \cite{Adcox:2003nr} 
can be also naturally explained because we have got right 
predictions for protons.

\section{Multiplicities of $\Theta ^+$ and $\Theta ^{*++}$}
\label{fivequark}

In this section we will give an estimate for 
multiplicities of $\Theta ^+$ and $\Theta ^{*++}$ at 130 AGeV and 200 AGeV. 
Several groups have already predicted the yields of $\Theta ^+$ 
in central Au+Au collisions by the statistical and coalescence models
\cite{Randrup:2003fq,Letessier:2003by,Chen:2003tn}.

In the QCM it is quite easy to estimate the production rate of
$\Theta ^+$ and $\Theta ^{*++}$.
The probabilities $P(uudd\ovl{s})$ and $P(uuud\ovl{s})$ for five quarks
$uudd\ovl{s}$ and $uuud\ovl{s}$ respectively to come together in rapidity 
can be easily obtained by the QCM. 
The overlapping of phase space (here rapidity) 
has been encoded in the probability $P$. 
Then the probabilities to form the pentaquark
$\Theta ^+$ and $\Theta ^{*++}$ can be
estimated by the spin counting: a system composed of five quarks,
each of which has spin-1/2, has totally $2^5$ spin states composed
of one spin-5/2, four spin-3/2 and five spin-1/2 states.
Assuming that all spin-1/2 states form $\Theta ^+$, the production rate for
$\Theta ^+$ is then $P(\Theta ^+)=\frac{10}{32}P(uudd\ovl{s})$.
If all spin-3/2 states go to $\Theta ^{*++}$, the
production rate of $\Theta ^{*++}$ is then
$P(\Theta ^{*++})=\frac{16}{32}P(uuud\ovl{s})$.
Note that the current spin counting method only 
provides a kind of upper limits for the yields. 
In the real calculation one must evaluate the transition probability  
which involves overlapping of the wave functions of 
pentaquark states and their quark constituents. 
Here we simplify the problem by assuming the transition amplitude as unity.
Also we neglect the contribution from the hadron-hadron 
rescattering in hadronic phase and assume the dominant source is from
quark combination. Our estimates for 
$\Theta ^+$ and $\Theta ^{*++}$ are shown in Table VI.
In summary of our results, at 130 AGeV there are 
about 1.16 $\Theta ^+$ and 1.86 $\Theta ^{*++}$ per 
rapidity produced in mid-rapidity in central Au+Au
collisions, while at 200 AGeV, the yields of $\Theta ^+$ and $\Theta ^{*++}$
are 1.3 and 2.08 per rapidity respectively. Our predictions agree with 
those given by the statistical models \cite{Randrup:2003fq,Letessier:2003by}. 
For comparison, the yield of $\Theta ^+$ in proton-proton
collision is predicted to be about $10^{-2}-10^{-3}$
\cite{Liu:2004ca,Bleicher:2004is}.

\section{Multiplicity of $(\Omega\Omega)_{0^{+}}$}
\label{diomega}

We have calculated the ratios of strange anti-baryons to their
baryon counterparts and that of $\Omega$ to negative charged
particles. The results are found to be consistent with available
data, see Table IV-V.
Based on the result for $\Omega$,
we will predict in this section multiplicities of 
the di-baryon $(\Omega\Omega)_{0^{+}}$
in Au+Au collisions at 130 and 200 AGeV.

First we estimate the yield of $\Omega$ ($\ovl{\Omega}^{+}+
\Omega^{-}$). Assuming the ratio of $\Omega$ to $h^{-}$ is
constant for all rapidities, we get the yield of $\Omega$ at about
5 and 6 per event from the experimental data $h^{-}\approx 1/2
N^{data}_{ch} \approx 2050$ at 130 AGeV \cite{Back:2002wb} and
$h^{-}\approx 1/2 N^{data}_{ch} \approx 2450$ at 200 AGeV
\cite{Nouicer:2002ks} respectively. In the quark combination
picture, the production weight for $\Omega^{-}$ is proportional to
$\lambda_s^{3}$. 
We find that the multiplicities increase strongly with growing
$\lambda _s$, from 0.76 at $\lambda _s=0.3$ to 6.00 at $\lambda _s=0.7$. 
Our predictions for the $\Omega^{-}$ yields are listed in Table VI.
One sees that the yield of $\Omega^{-}$ per event is
about 2.74. Given the result for $\Omega^{-}$, we are
now in a position to estimate the multiplicity of
$(\Omega\Omega)_{0^{+}}$. First we get the probability $P(ssssss)$
for six s-quarks to come together under the rule of our QCM 
and then we apply the spin-counting to estimate 
the yield of $(\Omega\Omega)_{0^{+}}$.
There are totally $2^6$ spin states for a six-quarks system. In
terms of total spin, these are one spin-3, five spin-2, nine
spin-1 and five spin-0 states. We assume that all spin-0 states go
to $(\Omega\Omega)_{0^{+}}$. The yield of $(\Omega\Omega)_{0^{+}}$
is then $\frac{5}{64}$ times that of six $s$ quarks. 
The results are shown in Table VI.
Same as the results for pentaquarks, the yields of 
$(\Omega\Omega)_{0^{+}}$ in Table VI are only regarded as 
a kind of upper limits because we have neglected the effect of 
the transition amplitude for 6 s-quarks to form a di-omega. 
One sees that the multiplicities 
of $(\Omega\Omega)_{0^+}$ are about $1.86\times 10^{-4}$ 
at 130 AGeV and $2.52\times 10^{-4}$ at
200 AGeV in an event respectively. 
Our predictions agree in magnitude with that
of the coalescence models \cite{Pal:2001rp} and are 
within the scope of present RHIC experiments. 
Normally one can detect $(\Omega\Omega)_{0^+}$ through its 
decay product. Because the binding energy of 
$(\Omega\Omega)_{0^+}$ is relatively large which prevent it from 
strong decay, $(\Omega\Omega)_{0^+}$ mainly 
weakly decays into one $\Omega$ plus the decay products 
of $\Omega$ (three or more body decay) or into 
$\Omega$ and $\Xi$ (two body decay). 
If it is produced in the experiments, one could 
detect it through the above decay channels. 
For other experimental traces of dibaryons, see, for example, 
Ref. \cite{Schaffner-Bielich:1999sy,Schaffner-Bielich:1996eh}.

\section{Summary}
\label{summ}

In this paper,  we extend our quark combination model, which
is very successful in describing the hadron production
in electron-positron and proton-proton(anti-proton) collisions,
to reproduce the multiplicity data in heavy ion collisions at RHIC 
energies 130 and 200 AGeV. The model can describe available data 
for transverse momentum spectra for pions, kaons and protons,  
especially the $p_T$ scaling behavior can be well reproduced 
within our simple model. It can explain the anormaly of 
the proton to pion ratio at intermediate transverse momenta.
A variety of ratios for anti-baryons to baryons, 
single-strange hadrons to non-strange ones, and double-strange hadrons
to single-strange ones can be all reproduced in our QCM.
With the approximation that the transition probability  
for constituent quarks to form an exotic state is taken as unity if 
they come togather in rapidity according to our combination rule, 
we finally estimate the yields of $\Theta ^+$, 
$\Theta ^{*++}$ and $(\Omega\Omega)_{0^{+}}$
in Au+Au collisions at 130 and 200 AGeV.
These yields can be regarded as a kind of upper limits. 
The multiplicities of $\Theta ^+$ and $\Theta ^{*++}$ are
estimated to be of order 1 and 2 per rapidity 
in mid-rapidity in a central event respectively, 
while the rate of $(\Omega\Omega)_{0^{+}}$ 
is of the magnitude $10^{-5}$ per rapidity. 
The multiplicites of $\Theta ^+$ and $\Theta ^{*++}$ 
are found to be almost independent of the strangeness 
suppression factor $\lambda_s$, while that of $(\Omega\Omega)_{0^{+}}$
increases strongly with growing $\lambda_s$.

\subsection*{Acknowledgments}

The authors thank C.R. Ching, S.-Y. Li, Z.-T. Liang,
T.H. Ho, Z.-G. Si, F. Wang, Y.W. Yu and Z.Y. Zhang for helpful discussions.
Q.W. thanks K. Redlich for critically read the manuscript and 
for many insightful comments. 
Q.W. acknowledges support by the Virtual Institute VH-VI-041
of the Helmholtz Association of National Research Centers.
The work of F.L.S. and Q.B.X. is supported in part by the
National Natural Science Foundation of China under
grant 10475049 and University Doctor Point Foundation of china 
under grant 20030422064.


\begin{thebibliography}{999}



\bibitem{Back:2002wb}
B.~B.~Back {\it et al.},
Phys.\ Rev.\ Lett.\  {\bf 91}, 052303 (2003)
[arXiv:nucl-ex/0210015].

\bibitem{Adler:2002sw}
C.~Adler {\it et al.}  [STAR Collaboration],
Phys.\ Rev.\ C {\bf 66}, 061901 (2002)
[arXiv:nucl-ex/0205015].


\bibitem{Adams:2002pf}
J.~Adams {\it et al.}  [STAR Collaboration],
Phys.\ Lett.\ B {\bf 567}, 167 (2003)
[arXiv:nucl-ex/0211024].

\bibitem{Hippolyte:2003yf}
B.~Hippolyte  [STAR Collaboration],
arXiv:nucl-ex/0306017.



\bibitem{Adler:2002wn}
C.~Adler {\it et al.}  [STAR Collaboration],
Phys.\ Lett.\ B {\bf 595}, 143 (2004)
[arXiv:nucl-ex/0206008.]


\bibitem{Adcox:2003nr}
K.~Adcox {\it et al.}  [PHENIX Collaboration],
Phys.\ Rev.\ C {\bf 69}, 024904 (2004)
[arXiv:nucl-ex/0307010].

\bibitem{Back:2002ks}
B.~B.~Back {\it et al.}  [PHOBOS Collaboration],
Phys.\ Rev.\ C {\bf 67}, 021901 (2003)
[arXiv:nucl-ex/0206012].


\bibitem{Filimonov:2003xu}
K.~Filimonov  [STAR Collaboration],
arXiv:hep-ex/0306056.





\bibitem{Braun-Munzinger:2003zd}
P.~Braun-Munzinger, K.~Redlich and J.~Stachel,
arXiv:nucl-th/0304013.

\bibitem{Braun-Munzinger:2003zz}
P.~Braun-Munzinger, J.~Stachel and C.~Wetterich,
arXiv:nucl-th/0311005.



\bibitem{Braun-Munzinger:1999qy}
P.~Braun-Munzinger, I.~Heppe and J.~Stachel,
Phys.\ Lett.\ B {\bf 465}, 15 (1999)
[arXiv:nucl-th/9903010].

\bibitem{Braun-Munzinger:2001ip}
P.~Braun-Munzinger, D.~Magestro, K.~Redlich and J.~Stachel,
Phys.\ Lett.\ B {\bf 518}, 41 (2001)
[arXiv:hep-ph/0105229].

\bibitem{Cleymans:1999st}
J.~Cleymans and K.~Redlich,
Phys.\ Rev.\ C {\bf 60}, 054908 (1999)
[arXiv:nucl-th/9903063].

\bibitem{Rafelski:pu}
J.~Rafelski and B.~Muller,
Phys.\ Rev.\ Lett.\  {\bf 48}, 1066 (1982)
[Erratum-ibid.\  {\bf 56}, 2334 (1986)].

\bibitem{Muller:1980kf}
B.~Muller and J.~Rafelski,
Phys.\ Lett.\ B {\bf 101}, 111 (1981).


\bibitem{Koch:1986ud}
P.~Koch, B.~Muller and J.~Rafelski,
Phys.\ Rept.\  {\bf 142}, 167 (1986).


\bibitem{Torrieri:2004zz}
G.~Torrieri, W.~Broniowski, W.~Florkowski, J.~Letessier and J.~Rafelski,
arXiv:nucl-th/0404083.

\bibitem{Torrieri:2003nh}
G.~Torrieri and J.~Rafelski,
J.\ Phys.\ G {\bf 30}, S557 (2004)
[arXiv:nucl-th/0305071].

\bibitem{Bratkovskaya:2000qy}
E.~L.~Bratkovskaya, W.~Cassing, C.~Greiner, M.~Effenberger,
U.~Mosel and A.~Sibirtsev,
Nucl.\ Phys.\ A {\bf 675}, 661 (2000)
[arXiv:nucl-th/0001008].

\bibitem{Schenke:2003mj}
B.~Schenke and C.~Greiner,
J.\ Phys.\ G {\bf 30}, 597 (2004)
[arXiv:nucl-th/0305008].

\bibitem{Rischke:2001bt}
D.~H.~Rischke,
Nucl.\ Phys.\ A {\bf 698}, 153 (2002)
[arXiv:nucl-th/0104071].




\bibitem{Hwa:2002tu}
R.~C.~Hwa and C.~B.~Yang,
Phys.\ Rev.\ C {\bf 67}, 034902 (2003)
[arXiv:nucl-th/0211010].

\bibitem{Hwa:2004ng}
R.~C.~Hwa and C.~B.~Yang,
Phys.\ Rev.\ C {\bf 70}, 024905 (2004)
[arXiv:nucl-th/0401001.]


\bibitem{Greco:2003xt}
V.~Greco, C.~M.~Ko and P.~Levai,
Phys.\ Rev.\ Lett.\  {\bf 90}, 202302 (2003)
[arXiv:nucl-th/0301093].

\bibitem{Greco:2004rm}
V.~Greco and C.~M.~Ko,
arXiv:nucl-th/0405040.

\bibitem{Greco:2003mm}
V.~Greco, C.~M.~Ko and P.~Levai,
Phys.\ Rev.\ C {\bf 68}, 034904 (2003)
[arXiv:nucl-th/0305024].


\bibitem{Fries:2003vb}
R.~J.~Fries, B.~Muller, C.~Nonaka and S.~A.~Bass,
Phys.\ Rev.\ Lett.\  {\bf 90}, 202303 (2003)
[arXiv:nucl-th/0301087].

\bibitem{Nonaka:2003hx}
C.~Nonaka, R.~J.~Fries and S.~A.~Bass,
Phys.\ Lett.\ B {\bf 583}, 73 (2004)
[arXiv:nucl-th/0308051].


\bibitem{Fries:2003kq}
R.~J.~Fries, B.~Muller, C.~Nonaka and S.~A.~Bass,
Phys.\ Rev.\ C {\bf 68}, 044902 (2003)
[arXiv:nucl-th/0306027].




\bibitem{Adcox:2001jp}
K.~Adcox {\it et al.}  [PHENIX Collaboration],
Phys.\ Rev.\ Lett.\  {\bf 88}, 022301 (2002)
[arXiv:nucl-ex/0109003].

\bibitem{Adler:2002tq}
C.~Adler {\it et al.}  [STAR Collaboration],
Phys.\ Rev.\ Lett.\  {\bf 90}, 082302 (2003)
[arXiv:nucl-ex/0210033].

\bibitem{Adcox:2001mf}
K.~Adcox {\it et al.}  [PHENIX Collaboration],
Phys.\ Rev.\ Lett.\  {\bf 88}, 242301 (2002)
[arXiv:nucl-ex/0112006].


\bibitem{Adler:2001bp}
C.~Adler {\it et al.}  [the STAR Collaboration],
Phys.\ Rev.\ Lett.\  {\bf 86}, 4778 (2001)
[Erratum-ibid.\  {\bf 90}, 119903 (2003)]
[arXiv:nucl-ex/0104022].

\bibitem{Adler:2002pba}
C.~Adler {\it et al.}  [STAR Collaboration],
Phys.\ Rev.\ Lett.\  {\bf 89}, 092301 (2002)
[arXiv:nucl-ex/0203016].

\bibitem{Adcox:2002au}
K.~Adcox {\it et al.}  [PHENIX Collaboration],
Phys.\ Rev.\ Lett.\  {\bf 89}, 092302 (2002)
[arXiv:nucl-ex/0204007].


\bibitem{Heinz:2004pj}
U.~W.~Heinz,
arXiv:nucl-th/0407067.



















\bibitem{Anisovich:1972pq}
V.~V.~Anisovich and V.~M.~Shekhter,
Nucl.\ Phys.\ B {\bf 55}, 455 (1973).


\bibitem{Bjorken:1973mh}
J.~D.~Bjorken and G.~R.~Farrar,
Phys.\ Rev.\ D {\bf 9}, 1449 (1974).


\bibitem{Andersson:ia}
B.~Andersson, G.~Gustafson, G.~Ingelman and T.~Sjostrand,
Phys.\ Rept.\  {\bf 97}, 31 (1983).


\bibitem{Webber:1983if}
B.~R.~Webber,
Nucl.\ Phys.\ B {\bf 238}, 492 (1984).

\bibitem{Winter:2003tt}
J.~C.~Winter, F.~Krauss and G.~Soff,
Eur.\ Phys.\ J.\ C {\bf 36}, 381 (2004)
[arXiv:hep-ph/0311085.]





\bibitem{Xie:wi}
Q.~B.~Xie and X.~M.~Liu,
Phys.\ Rev.\ D {\bf 38}, 2169 (1988).


\bibitem{Xie:ap}
Q.~B.~Xie,
Proceedings of the 19th International Symposium on Multiparticle
Dynamics, Arles, France,1988, edited by D.Schiff and J.Tran Thanh
Vann (world scitific,1988)p369.


\bibitem{Liang:ya}
Z.~T.~Liang and Q.~B.~Xie,
Phys.\ Rev.\ D {\bf 43}, 751 (1991).

\bibitem{Wang:gx}
Q.~Wang and Q.~B.~Xie,
Phys.\ Rev.\ D {\bf 52}, 1469 (1995).

\bibitem{Wang:ch}
Q.~Wang and Q.~B.~Xie,
J.\ Phys.\ G {\bf 21}, 897 (1995).

\bibitem{Zhao:hq}
J.~Q.~Zhao, Q.~Wang and Q.~B.~Xie,
Sci.\ Sin.\ A {\bf 38}, 1474 (1995).

\bibitem{Wang:dg}
Q.~Wang, X.~M.~Liu and Q.~B.~Xie,
High Energy Phys.\ Nucl.\ Phys.\  {\bf 19}, 281 (1995).

\bibitem{Wang:jy}
Q.~Wang, Z.~G.~Si and Q.~B.~Xie,
Int.\ J.\ Mod.\ Phys.\ A {\bf 11}, 5203 (1996).

\bibitem{Wang:pg}
Q.~Wang, Q.~B.~Xie and Z.~G.~Si,
Phys.\ Lett.\ B {\bf 388}, 346 (1996).

\bibitem{Si:ux}
Z.~G.~Si, Q.~B.~Xie and Q.~Wang,
Commun.\ Theor.\ Phys.\  {\bf 28}, 85 (1997).


\bibitem{Si:1999}
Z.~G.~Si, Q.~B.~Xie , High Energy Phys.\ Nucl.\ Phys.\  {\bf 23},
445 (1999).



\bibitem{Wang:1999xz}
Q.~Wang, G.~Gustafson and Q.~B.~Xie,
Phys.\ Rev.\ D {\bf 62}, 054004 (2000)
[arXiv:hep-ph/9912310].

\bibitem{Wang:2000bv}
Q.~Wang, G.~Gustafson, Y.~Jin and Q.~B.~Xie,
Phys.\ Rev.\ D {\bf 64}, 012006 (2001)
[arXiv:hep-ph/0011362].

\bibitem{Li:2002eq}
S.~Y.~Li, F.~l.~Shao, Q.~B.~Xie and Q.~Wang,
Phys.\ Rev.\ D {\bf 65}, 077503 (2002)
[arXiv:hep-ph/0201169].

\bibitem{Shao:2003ir}
F.~L.~Shao, Q.~B.~Xie, S.~Y.~Li and Q.~Wang,
Phys.\ Rev.\ D {\bf 69}, 054007 (2004)
[arXiv:hep-ph/0306090].
























\bibitem{Jaffe:1976ii}
R.~L.~Jaffe,
SLAC-PUB-1774
{\it Talk presented at the Topical Conf. on Baryon Resonances,
Oxford, Eng., Jul 5-9, 1976}.

\bibitem{Strottman:qu}
D.~Strottman,
Phys.\ Rev.\ D {\bf 20}, 748 (1979).


\bibitem{Diakonov:1997mm}
D.~Diakonov, V.~Petrov and M.~V.~Polyakov,
Z.\ Phys.\ A {\bf 359}, 305 (1997)
[arXiv:hep-ph/9703373].


\bibitem{Wu:2003xc}
B.~Wu and B.~Q.~Ma,
Phys.\ Lett.\ B {\bf 586}, 62 (2004)
[arXiv:hep-ph/0312326].

\bibitem{Wu:2003mc}
B.~Wu and B.~Q.~Ma,
Phys.\ Rev.\ D {\bf 69}, 077501 (2004)
[arXiv:hep-ph/0312041].

\bibitem{Liu:2004qx}
Y.~R.~Liu, A.~Zhang, P.~Z.~Huang, W.~Z.~Deng, X.~L.~Chen and S.~L.~Zhu,
Phys.\ Rev.\ D {\bf 70}, 094045 (2004)
[arXiv:hep-ph/0404123.]




\bibitem{Nakano:2003qx}
T.~Nakano {\it et al.}  [LEPS Collaboration],
Phys.\ Rev.\ Lett.\  {\bf 91}, 012002 (2003)
[arXiv:hep-ex/0301020].





\bibitem{Barmin:2003vv}
V.~V.~Barmin {\it et al.}  [DIANA Collaboration],
Phys.\ Atom.\ Nucl.\  {\bf 66}, 1715 (2003)
[Yad.\ Fiz.\  {\bf 66}, 1763 (2003)]
[arXiv:hep-ex/0304040].




\bibitem{Stepanyan:2003qr}
S.~Stepanyan {\it et al.}  [CLAS Collaboration],
Phys.\ Rev.\ Lett.\  {\bf 91}, 252001 (2003)
[arXiv:hep-ex/0307018].


\bibitem{Airapetian:2003ri}
A.~Airapetian {\it et al.}  [HERMES Collaboration],
Phys.\ Lett.\ B {\bf 585}, 213 (2004), [arXiv:hep-ex/0312044].

\bibitem{Kubarovsky:2003fi}
V.~Kubarovsky {\it et al.}  [CLAS Collaboration],
Phys.\ Rev.\ Lett.\  {\bf 92}, 032001 (2004)
[Erratum-ibid.\  {\bf 92}, 049902 (2004)]
[arXiv:hep-ex/0311046].

\bibitem{Aubert:2004ps}
B.~Aubert {\it et al.}  [BABAR Collaboration],
arXiv:hep-ex/0408037.




\bibitem{Jaffe:1977cv}
R.~L.~Jaffe,
Phys.\ Rev.\ D {\bf 17}, 1444 (1978).


\bibitem{Jaffe:1976yi}
R.~L.~Jaffe,
Phys.\ Rev.\ Lett.\  {\bf 38}, 195 (1977)
[Erratum-ibid.\  {\bf 38}, 617 (1977)].


\bibitem{Buchmann:wy}
A.~J.~Buchmann, G.~Wagner, K.~Tsushima, L.~Y.~Glozman and A.~Faessler,
Prog.\ Part.\ Nucl.\ Phys.\  {\bf 36}, 383 (1996).

\bibitem{Buchmann:1998mi}
A.~J.~Buchmann, G.~Wagner and A.~Faessler,
Phys.\ Rev.\ C {\bf 57}, 3340 (1998)
[arXiv:nucl-th/9803025].

\bibitem{Wang:2004nv}
F.~Wang, J.~l.~Ping, D.~Qing and T.~Goldman,
arXiv:nucl-th/0406036.

\bibitem{Pang:2004mm}
H.~R.~Pang, J.~l.~Ping, L.~Z.~Chen, F.~Wang and T.~Goldman,
Phys.\ Rev.\ C {\bf 70}, 035201 (2004)
[arXiv:hep-ph/0406145.]






\bibitem{Zhang:ju}
Z.~Y.~Zhang, Y.~W.~Yu and X.~Q.~Yuan,
Nucl.\ Phys.\ A {\bf 670}, 178 (2000).

\bibitem{Li:2000cb}
Q.~B.~Li, P.~N.~Shen, Z.~Y.~Zhang and Y.~W.~Yu,
Nucl.\ Phys.\ A {\bf 683}, 487 (2001)
[arXiv:nucl-th/0009038].

\bibitem{Yu:2002jm}
Y.~W.~Yu, P.~Wang, Z.~Y.~Zhang, C.~R.~Ching, T.~H.~Ho and L.~Y.~Chu,
Phys.\ Rev.\ C {\bf 66}, 015205 (2002).


\bibitem{Zhang:ny}
Z.~Y.~Zhang, Y.~W.~Yu, P.~N.~Shen, L.~R.~Dai, A.~Faessler and U.~Straub,
Nucl.\ Phys.\ A {\bf 625}, 59 (1997).


\bibitem{Pal:2001rp}
S.~Pal, C.~M.~Ko and Z.~Y.~Zhang,
arXiv:nucl-th/0107070.


\bibitem{Zhang:sv}
Z.~Y.~Zhang, Y.~W.~Yu, C.~R.~Ching, T.~H.~Ho and Z.~D.~Lu,
Phys.\ Rev.\ C {\bf 61}, 065204 (2000).



\bibitem{Nouicer:2002ks}
R.~Nouicer {\it et al.}  [PHOBOS Collaboration],
arXiv:nucl-ex/0208003.





\bibitem{Hofmann:1999jx}
M.~Hofmann, M.~Bleicher, S.~Scherer, L.~Neise, H.~Stocker and W.~Greiner,
Phys.\ Lett.\ B {\bf 478}, 161 (2000)
[arXiv:nucl-th/9908030].

\bibitem{Hofmann:1988gy}
W.~Hofmann,
Ann.\ Rev.\ Nucl.\ Part.\ Sci.\  {\bf 38}, 279 (1988).

\bibitem{Casher:wy}
A.~Casher, H.~Neuberger and S.~Nussinov,
Phys.\ Rev.\ D {\bf 20}, 179 (1979).

\bibitem{Casher:gw}
A.~Casher, H.~Neuberger and S.~Nussinov,
Phys.\ Rev.\ D {\bf 21}, 1966 (1980).

\bibitem{Scheck}
H.\ Scheck, Nucl.\ Phys.\ B (Proc. Suppl.) {\bf 1}, 291(1988).


\bibitem{Adler:2003qi}
S.~S.~Adler {\it et al.}  [PHENIX Collaboration],
Phys.\ Rev.\ Lett.\  {\bf 91}, 072301 (2003)
[arXiv:nucl-ex/0304022].

\bibitem{Hwa:2003bn}
R.~C.~Hwa and C.~B.~Yang,
Phys.\ Rev.\ C {\bf 67}, 064902 (2003)
[arXiv:nucl-th/0302006].


\bibitem{Randrup:2003fq}
J.~Randrup,
Phys.\ Rev.\ C {\bf 68}, 031903 (2003)
[arXiv:nucl-th/0307042].

\bibitem{Letessier:2003by}
J.~Letessier, G.~Torrieri, S.~Steinke and J.~Rafelski,
Phys.\ Rev.\ C {\bf 68}, 061901 (2003)
[arXiv:hep-ph/0310188].

\bibitem{Chen:2003tn}
L.~W.~Chen, V.~Greco, C.~M.~Ko, S.~H.~Lee and W.~Liu,
Phys.\ Lett.\ B {\bf 601}, 34 (2004) 
[arXiv:nucl-th/0308006].







\bibitem{Liu:2004ca}
F.~M.~Liu, H.~Stoecker and K.~Werner,
Phys.\ Lett.\ B {\bf 597}, 333 (2004)
[arXiv:hep-ph/0404156.]


\bibitem{Bleicher:2004is}
M.~Bleicher, F.~M.~Liu, J.~Aichelin, T.~Pierog and K.~Werner,
Phys.\ Lett.\ B {\bf 595}, 288 (2004)
arXiv:hep-ph/0401049.


\bibitem{Schaffner-Bielich:1999sy}
J.~Schaffner-Bielich, R.~Mattiello and H.~Sorge,
Phys.\ Rev.\ Lett.\  {\bf 84}, 4305 (2000)
[arXiv:nucl-th/9908043].

\bibitem{Schaffner-Bielich:1996eh}
J.~Schaffner-Bielich, C.~Greiner, A.~Diener and H.~Stocker,
Phys.\ Rev.\ C {\bf 55}, 3038 (1997)
[arXiv:nucl-th/9611052].





\end{thebibliography}
\end{document}